%% file: egpaper_final.tex
\documentclass[10pt,twocolumn,letterpaper]{article}

\usepackage{iccv}
\usepackage{times}
\usepackage{epsfig}
\usepackage{graphicx}
\usepackage{amsmath}
\usepackage{amssymb}

\usepackage{array}
\usepackage{lipsum}
\usepackage{breqn}
\usepackage{graphicx}
\usepackage{amsmath}
\usepackage{amssymb}
\usepackage{booktabs}
\usepackage{multirow}
\usepackage{multicol}
\usepackage{tikz}

\def\checkmark{\tikz\fill[scale=0.4](0,.35) -- (.25,0) -- (1,.7) -- (.25,.15) -- cycle;} 

\usepackage[pagebackref=true,breaklinks=true,letterpaper=true,colorlinks,bookmarks=false]{hyperref}

\iccvfinalcopy 

\def\iccvPaperID{12426} 

\ificcvfinal\fi

\begin{document}

\title{Mirror U-Net: Marrying Multimodal Fission with Multi-task Learning \\ for Semantic Segmentation in Medical Imaging}
\author{$^{1,3}$Zdravko Marinov$^*$ \hspace{0.2cm} $^1$Simon Reiß$^*$   \hspace{0.2cm}  $^{2,5,6}$David Kersting$^{\dagger}$ \hspace{0.2cm} $^{4,7}$Jens Kleesiek$^{\dagger}$ \hspace{0.2cm}   $^1$Rainer Stiefelhagen$^*$ \\
\small$^1$Institute for Anthropomatics and Robotics (IAR), Karlsruhe Institute of Technology, \small$^2$German Cancer Consortium (DKTK) \\ 
\small$^3$HIDSS4Health - Helmholtz Information and Data Science School for Health  \\ 
\small$^4$Institute for AI in Medicine (IKIM), University Hospital Essen, \small$^5$Department of Nuclear Medicine, University Hospital Essen \\ 
\small$^6$West German Cancer Center, \small$^7$Cancer Research Center Cologne Essen (CCCE), University Medicine Essen \\
{\tt\small $^*$firstname.lastname@kit.edu} \\
\tt\small $^{\dagger}$firstname.lastname@uk-essen.de}

\maketitle
\ificcvfinal\thispagestyle{empty}\fi

\input{sections/abstract}

\input{sections/introduction}

\input{sections/related_work}

\input{sections/method}

\input{sections/experiments}

\input{sections/conclusion}

{\small
\bibliographystyle{ieee_fullname}
\bibliography{egbib}
}

\end{document}

%% file: sections/abstract.tex
\begin{abstract}

Positron Emission Tomography (PET) and Computed Tomography (CT) are routinely used together to detect tumors. PET/CT segmentation models can automate tumor delineation, however, current multimodal models do not fully exploit the complementary information in each modality, as they either concatenate PET and CT data or fuse them at the decision level. To combat this, we propose Mirror U-Net, which replaces traditional fusion methods with multimodal fission by factorizing the multimodal representation into modality-specific decoder branches and an auxiliary multimodal decoder. At these branches, Mirror U-Net assigns a task tailored to each modality to reinforce unimodal features while preserving multimodal features in the shared representation. In contrast to previous methods that use either fission or multi-task learning, Mirror U-Net combines both paradigms in a unified framework. We explore various task combinations and examine which parameters to share in the model. We evaluate Mirror U-Net on the AutoPET PET/CT and on the multimodal MSD BrainTumor datasets, demonstrating its effectiveness in multimodal segmentation and achieving state-of-the-art performance on both datasets. Our code will be made publicly available.


\end{abstract}

%% file: sections/introduction.tex
\section{Introduction}
\label{sec:intro}
\begin{figure}[t]
    \centering
    \includegraphics[scale=1.2]{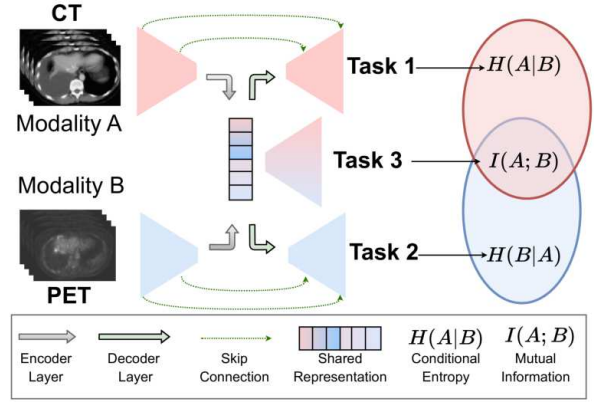}
    \caption{Mirror U-Net combines multimodal fission \cite{liang2022foundations} and multi-task learning. We obtain a shared representation for both modalities and feed it into modality-specific decoders, each optimized for a tailored task to learn useful features from its modality. Tasks 1 and 2 use modality-specific features via the skip connections but are also conditioned by the other modality via the shared representation, hence focusing on the conditional entropy between the modalities. Task 3, on the other hand, only processes the shared representation, focusing on the mutual information.}
    \label{fig:teaser}
\end{figure}
PET/CT scans are commonly used for cancer diagnosis and therapy to estimate tumor characteristics, such as size, location, and changes over time \cite{yap2004image}. PET data can highlight areas with a high metabolic activity, which is typical for tumors \cite{ben200918f}, by administering a radioactive tracer like Fluorodeoxyglucose (FDG). To provide detailed anatomical information and aid in accurate tumor localization, CT scans are typically used in conjunction with PET scans \cite{townsend2004pet}.

Deep learning models can automatically segment lesions in PET/CT scans, providing radiologists with metabolic tumor volume (MTV) and shape information as biomarkers to monitor disease progression \cite{gatidis2022whole, isensee2017brain}. However, high metabolic activity in PET is not specific to tumors and can be found in organs and regions with inflammation or infection \cite{bi2014multi}. Additionally, while CT scans provide anatomical information, they are not sufficient for visualizing lesions on their own \cite{townsend2004pet}. These factors make tumor segmentation from PET/CT data challenging, especially due to the limited availability of voxel-wise labeled PET/CT datasets. As a result, current multimodal PET/CT segmentation models have yet to demonstrate reliability for clinical use.

Recently, the AutoPET MICCAI 2022 Challenge \cite{gatidis2022whole} released a large-scale labeled PET/CT database of 1014 studies involving 900 patients. However, the top-performing methods in the final leaderboard rely on either early fusion \cite{peng2022automatic, ye2022exploring, bendazzoli2022priornet, zhong2022autopet, heiliger2022autopet} and/or late fusion ensembles \cite{zhang2022whole, sibille2022whole,heiliger2022autopet}, which do not fully leverage the complementary information in the PET and CT modalities.

We propose Mirror U-Net, a unified framework that combines multimodal fission~\cite{liang2022foundations} and multi-task learning. Rather than fusing modality features, Mirror U-Net factorizes multimodal features into modality-specific decoder branches and an auxiliary multimodal decoder, allowing us to disentangle modality-specific features, such as metabolic and anatomical cues in PET/CT, from multimodal features. Using an encoder-decoder U-Net model~\cite{ronneberger2015u} for each modality, we obtain modality-specific features and share layers between them to produce the multimodal representation, as shown in Figure \ref{fig:teaser}. To emphasize the dichotomy between modalities and their shared features, we extend multimodal fission with multi-task learning, investigating four combinations of tasks tailored to unimodal or multimodal features. In our qualitative experiments, we demonstrate how Mirror U-Net utilizes the complementary information from each modality. Our approach surpasses traditional fusion schemes, as well as fission-only or multi-task-only methods, achieving state-of-the-art performance on two benchmarks, AutoPET~\cite{gatidis2022whole} and MSD BrainTumor~\cite{antonelli2022medical}.

Our contributions are summarized as follows:
\begin{enumerate}
\itemsep0em
    \item A novel unification of multimodal fission and multi-task learning for multimodal medical segmentation.
    \item Mirror U-Net - a simple yet powerful multimodal fission architecture that achieves state-of-the-art performance on AutoPET~\cite{gatidis2022whole} and MSD BrainTumor~\cite{antonelli2022medical}, demonstrating great potential for deploying PET/CT segmentation models in clinical practice.
    \item We conduct extensive experiments to determine which tasks to assign to the decoder branches and which layers to share to obtain the multimodal representation.
\end{enumerate}

%% file: sections/related_work.tex
\section{Related Work}
\label{sec:related_work}

\subsection{Multimodal Fusion}
Combining multiple imaging modalities for automatic segmentation has advanced significantly, particularly in PET/CT \cite{xue2021multi, bourigault2021multimodal, fu2021multimodal, gatidis2022whole, hallitschke2023multimodal}, CT/MRI \cite{zhang2018translating, jiang2020self, jiang2019cross}, and multi-contrast MRI \cite{menze2014multimodal, isensee2017brain, jiang2019two}. Multimodal approaches often use early fusion, where modalities are concatenated into a single input \cite{isensee2017brain, jiang2019two, bourigault2021multimodal, peng2022automatic, ye2022exploring, bendazzoli2022priornet, zhong2022autopet, liu2022autopet, heiliger2022autopet}, or late fusion, where predictions from unimodal models are combined \cite{zhang2022whole, sibille2022whole, weisman2022automated}. However, late fusion may not exploit the mutual information in cross-modal representations, and early fusion may not highlight the contribution of each modality to the task \cite{marinov2022modselect}. Fu \etal \cite{fu2021multimodal} use a PET U-Net model \cite{ronneberger2015u} to guide a CT model in a cascade framework, while Xue \etal \cite{xue2021multi} segment liver lesions by combining predictions from low- and high-level feature maps in separate PET/CT decoders. Early fusion is shown to outperform late and middle fusion for brain tumor segmentation from MRI/PET/CT data \cite{guo2019deep}. Other approaches translate CT into MRI images via GANs with cycle- and shape-consistency \cite{zhang2018translating}, organ attention \cite{jiang2020self}, or using a cross-modality prior \cite{jiang2019cross} to generate more data. In contrast to fusion approaches, Mirror U-Net disentangles unimodal and multimodal features using modality-specific decoders and a multimodal decoder, allowing us to define tasks for each decoder to explicitly focus on the modality's strengths, such as anatomical knowledge from CT and metabolic activity in PET.

\subsection{Multimodal Fission}
Multimodal fission decomposes data into multimodal and unimodal information that captures the unique structure and semantics of each modality \cite{liang2022foundations,tsai2018learning,hsu2018disentangling}. This separation can be achieved through disentangled representation learning \cite{tsai2018learning,hsu2018disentangling} or explicit separation of unimodal and multimodal pathways in the model \cite{kendall2018multi,hickson2019floors,maninis2019attentive}. In our work, we adopt the latter using our Mirror U-Net architecture. Joze \etal \cite{joze2020mmtm} introduce a Multi-Modal Transfer Module as an independent multimodal pathway, while in \cite{shu2022expansion}, multimodal and unimodal paths are coordinated by squeeze and excitation operations. Valindria \etal \cite{valindria2018multi} use a similar architecture to Mirror U-Net but alternate training iterations between CT and MRI volumes for segmentation and do not employ multi-task learning. Mirror U-Net differs from existing fission approaches in that we combine it for the first time with multi-task learning to control the type of features learned in the modality-specific layers.

Hickson \etal \cite{hickson2022sharing} provide a summary of existing fission methods, noting that all previous methods either use a joint encoder before the fission into individual decoders \cite{liu2022sf, weninger2019multi, andrearczyk2021multi, meng2022deepmts} or share skip connections \cite{kuga2017multi}. In contrast, Mirror U-Net disentangles modalities by using modality-specific encoders and skip connections, both of which are essential to separate unimodal and multimodal features.

\subsection{Multi-Task Learning}
Multi-task learning has been widely used in medical image segmentation to exploit the correlation between different tasks \cite{andrearczyk2021multi, cheng2022fully, meng2022deepmts} or to regularize the segmentation \cite{mlynarski2019deep, weninger2019multi, liu2022sf}. For instance, Meng \etal~\cite{meng2022deepmts} train a survival prediction network on feature maps extracted from a PET/CT U-Net \cite{ronneberger2015u} and outperform the single-task model. Other approaches utilize image reconstruction as an auxiliary task to guide and regularize the segmentation \cite{weninger2019multi, liu2022sf}. Some methods employ the classification of tumor presence as an additional task to reduce false positives \cite{heiliger2022autopet} or prevent the network from learning irrelevant features \cite{mlynarski2019deep}. However, these multi-task methods are all limited to traditional multimodal fusion. In contrast, Mirror U-Net combines multimodal fission with multi-task learning, which has not been explored before. We show in our experiments that this combination outperforms multi-task fusion methods on two challenging datasets, despite using a simple architecture.

%% file: sections/method.tex
\begin{figure*}[t]
    \centering
    \includegraphics[width=\linewidth]{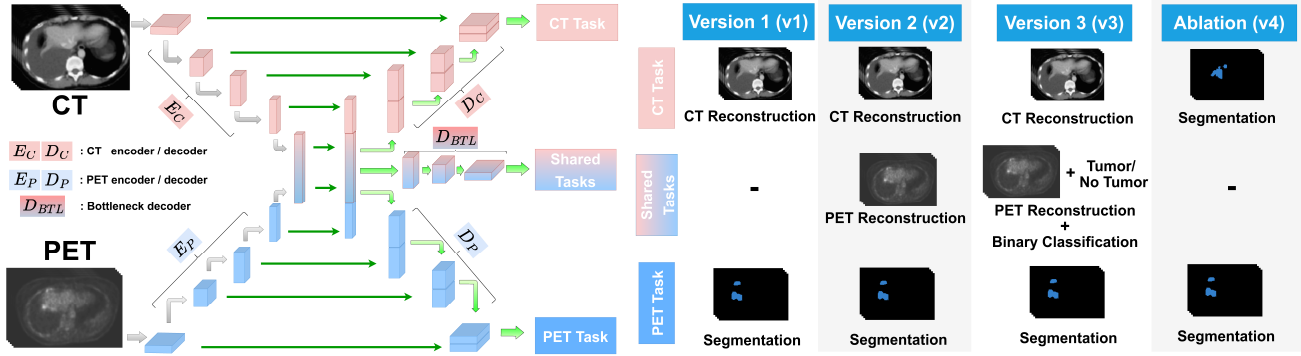}
    \caption{We explore 4 combinations of tasks \textbf{(v1) -- (v4)} by assigning different tasks to each decoder branch in Mirror U-Net. In \textbf{(v1)} we train the CT branch for CT reconstruction and assign the primary segmentation task to the PET branch. In \textbf{(v2)}, we extend \textbf{(v1)} by adding a bottleneck decoder that reconstructs PET data and in \textbf{(v3)} we also add a classifier, which predicts whether the patient is healthy or not. In \textbf{(v4)}, we train both the CT and PET branches for segmentation and fuse their predictions via a weighted sum of logits as an ablation study.}
    \label{fig:multi-task-settings}
    \vspace*{-0.2cm}
\end{figure*}
\section{Method}

\label{sec:method}
Mirror U-Net utilizes two 3D U-Net models \cite{ronneberger2015u, cciccek20163d} with shared bottleneck weights, resembling two U-Nets separated by a horizontal mirror. The first model processes CT data while the second model processes PET data, allowing the skip connections to the decoder branches to reinforce modality-specific features, while shared layers learn multimodal features. Liang \etal \cite{liang2022foundations} note that this approach expresses the conditional entropy of unimodal representations, indicating the additional information that each modality contributes to the task, whereas the shared representation expresses the mutual information. We explore 4 task combinations \textbf{(v1) -- (v4)}, visualized in Figure \ref{fig:multi-task-settings}, to showcase how various tasks leverage the disentangled information.

\subsection{Task Combinations}
\input{tables/notation}

We prioritize tumor segmentation as the primary task in our experiments and consider other tasks as auxiliary. We use a combination of Dice \cite{milletari2016v} and cross-entropy loss, denoted by $\mathcal{L}_{\text{DiceCE}}(y, \hat{y})$, where $y$ and $\hat{y}$ denote the ground-truth mask and model prediction, respectively. We use $E(\cdot)$ and $D(\cdot)$ to denote the encoder and decoder branches of Mirror U-Net, with $x$ representing the input PET/CT data. CT and PET branches and data are distinguished by subscripts $C$ and $P$, respectively, as summarized in Table \ref{tab:notation}. We explore four task combinations, labeled as \textbf{(v1) -- (v4)}, presented in Equations \ref{eqn:tf_eq} -- \ref{eqn:df_eq}, and depicted in Figure \ref{fig:multi-task-settings}.

\textbf{Version 1 (v1).} We explore the concept of transference \cite{liang2022foundations}, where knowledge from a source modality is transferred to a noisy target modality to adapt it for a primary task. Specifically, we transfer high-resolution anatomical knowledge from CT to its low-resolution PET counterpart via the multimodal representation. To achieve this, we train the CT branch for reconstruction via an $L_2$ loss, while training the PET branch for segmentation using Equation \ref{eqn:tf_eq}.

As lesions are often absent in CT and its high resolution is underutilized, a self-supervised task is better suited for the CT branch than segmentation. To prevent copying the CT input $x_C$ through skip connections, we use a transformation $\Phi(x_C)$, which can be Gaussian noise $G(x_C)$ or shuffled voxel patches $S(x_C)$ in the CT scan. The self-supervised $L_2$ loss and segmentation loss $\mathcal{L}_{\text{DiceCE}}$ are balanced using hyperparameters $\lambda_{\text{rec}}$, $\lambda_{\text{seg}}\in[0,1]$ in Equation \ref{eqn:tf_eq}. By encoding high-resolution anatomical information in the bottleneck via the CT reconstruction,  we enhance the primary segmentation performance for low-resolution PET scans.
\begin{align}
    \label{eqn:tf_eq}
    \begin{split}
    \mathcal{L}_{\text{V1}}(x, y) = &\lambda_{\text{rec}} \cdot ||x_C - D_C(E_C(\Phi(x_C)))||_2 + \\ + &\lambda_{\text{seg}} \cdot \mathcal{L}_{\text{DiceCE}}(y, D_P(E_P(x_P)))
    \end{split}
\end{align}

\textbf{Version 2 (v2).} To decouple modality-specific and multimodal features, we propose an additional bottleneck decoder, $D_{BTL}(\cdot)$, which is trained to reconstruct PET data, as shown in Equation \ref{eqn:mf_eq}. The reason we choose to reconstruct PET data is to condition the multimodal features to remain sensitive to regions with high metabolic uptake, a characteristic commonly associated with tumors \cite{ben200918f}. Our experiments show that such PET reconstruction provides spatial attention that guides the primary segmentation task toward active regions (see Figure \ref{fig:qualitative_results}). In contrast to modality-specific branches, which utilize skip connections, we omit them in the bottleneck to facilitate the decoupling of modality-specific and modality-shared information, differing from previous fission methods \cite{kuga2017multi}. Unlike \textbf{(v1)}, we do not apply any transformations $\Phi(\cdot)$ to $x_P$ since PET data cannot be copied via skip connections as there are none.
\begin{align}
    \label{eqn:mf_eq}
    \begin{split}
        \mathcal{L}_{\text{V2}}(x, y) &= \mathcal{L}_{\text{V1}}(x, y) +
        \\ + \lambda_{\text{rec}}\cdot||x_P - D_{BTL}(&E_C(x_C) \oplus E_P(x_P))||_2
    \end{split}
\end{align}

\textbf{Version 3 (v3).} Building on the \textbf{(v2)} model, we introduce a binary tumor classifier, $T(\cdot)$, to determine whether a tumor is present in the volume.  In AutoPET \cite{gatidis2022whole}, 513 of 1014 volumes are of tumor-free patients. Previous methods trained on AutoPET \cite{gatidis2022whole} have shown that training on all cases leads to conservative predictions for tumor-positive cases, resulting in under-segmentation \cite{heiliger2022autopet, liu2022autopet}. In contrast, training solely on positive cases fails to utilize the entire dataset and can lead to a high False-Positive-Volume (FPV) in healthy cases \cite{heiliger2022autopet, bendazzoli2022priornet, ye2022exploring}. To address these challenges, we propose a bottleneck classifier to regularize the network to predict empty masks in healthy cases and identify low-uptake tumors in positive cases. We extend the \textbf{(v2)} model with a classification task, as described in Equation \ref{eq:mfc}, where $\lambda_{\text{class}}\in[0,1]$ and $c\in\{0,1\}$ indicates tumor presence.
\begin{align}
    \label{eq:mfc}
    \begin{split}
      \mathcal{L}_{\text{V3}}(x, y) = \mathcal{L}_{\text{V2}}(x, y) &+
    \\  + \lambda_{\text{class}} \cdot \mathcal{L}_{\text{BCE}}(c, T(E_C(x_C) \oplus &E_P(x_P)))
    \end{split}
\end{align}

\textbf{Ablation (v4).} In our final version, we jointly train the modality-specific branches on segmentation by combining their logits through a weighted sum, as shown in Equation \ref{eqn:df_eq}. This is a challenging task for the CT branch as lesions are often not visually prominent in CT. To address this, we introduce a parameter $\theta \in [0.1, 0.5]$ to balance the CT- and PET-branch logits. A lower $\theta$ indicates a weaker reliance on CT data. We also explore the possibility of learning $\theta$ by the model, rather than manually tuning it. We refer to \textbf{(v4)} as an ablation since it is a single-task multimodal fission.
\begin{align}
\label{eqn:df_eq}
    \begin{split}
        \mathcal{L}_{\text{V4}}(x, y) &= \mathcal{L}_{\text{DiceCE}}(y, \hat{y})
    \\
        \hat{y} = (1 - \theta) \cdot D_P(E_P&(x_P)) + \theta \cdot D_C(E_C(x_C)))
    \end{split}
\end{align}

\subsection{Weight Sharing}
\begin{figure*}[t]
    \centering
    \includegraphics[scale=0.045]{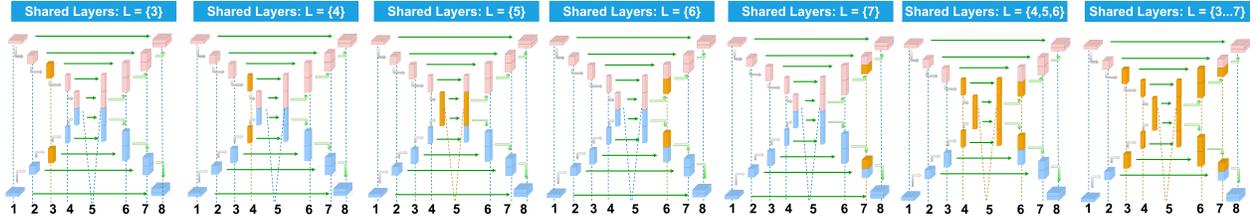}
    \caption{Different locations in the model to share the parameters between the modality-specific branches. Shared layers are colored orange. The set of layer indices $L \subset \{1...8\}$ are aligned with the layers via horizontal dashed lines. Best viewed in color.}
    \label{fig:weight_sharing}
\end{figure*}

While shared weights between the two U-Net models \cite{ronneberger2015u} facilitate learning a multimodal representation, the literature on the optimal location to share parameters within multimodal fission models is mostly confined to small ablation studies \cite{liang2022foundations, tsai2018learning}. To address this gap, we investigate different locations to share parameters between the branches (see Figure \ref{fig:weight_sharing}) and identify the optimal location based on empirical results. We index the layers of Mirror U-Net from 1 to 8 and denote the shared layers as $L$. We examine sharing layers before the bottleneck in the encoder branches where features are closer to the input modality, as well as sharing after the bottleneck in the decoders where features are more task-specific. We investigate how different shared layers impact the performance across the four versions \textbf{(v1) -- (v4)}, to determine if there is a consistent weight-sharing scheme that is optimal for all versions, and to evaluate the sensitivity of each version to changes in hyperparameters.

\subsection{Generalization to Brain Tumor Segmentation}
To demonstrate the generalizability of Mirror U-Net to other tasks and imaging modalities, we evaluate it on the MSD BrainTumor dataset \cite{antonelli2022medical}. It consists of 750 volumes of multimodal Magnetic Resonance Imaging (MRI) data, including (\textbf{T1}), post-contrast T1-weighted (\textbf{T1Gd}), (\textbf{T2}), and Fluid Attenuated Inversion Recovery (\textbf{FLAIR}) modalities. Similar to the complementary physiological and anatomical information in PET/CT data, we use the complementary FLAIR and T1Gd modalities, which are representative of the tumor edema (the accumulation of fluid around a tumor) and the tumor core (its central part), respectively. Together, the edema and core form the entire tumor. We use Mirror U-Net \textbf{(v2)} instead of \textbf{(v3)} as each volume of the dataset contains a tumor, and classification is not feasible. We replace the PET/CT branches with T1Gd/FLAIR and set the tasks for T1Gd and FLAIR to tumor core and edema segmentation, respectively. We also set the shared task to whole tumor segmentation. However, to obtain the final whole-tumor segmentation, we unite the predictions for core and edema from the T1Gd and FLAIR branches, since the whole tumor output from the bottleneck is used only as a regularization. We opt for segmentation tasks on all branches since unlike CT, FLAIR has a strong signal for edema segmentation. However, for consistency, we train a model \textbf{(v2)-rec} to reconstruct FLAIR and T1Gd in the FLAIR and shared branch, respectively, and segment all three tumor classes in the T1Gd branch, to demonstrate that Mirror U-Net can generalize well with the default tasks in Figure \ref{fig:multi-task-settings}.

\begin{figure}[b]
    \centering
    \includegraphics[scale=0.15]{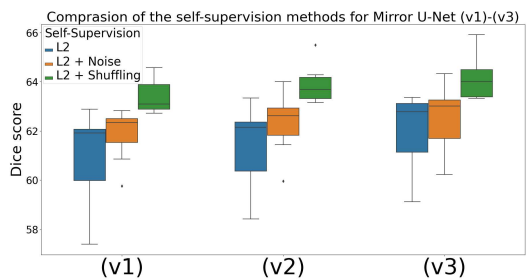}
    \caption{Comparison between the self-supervision methods used in Mirror U-Net \textbf{(v1)--(v3)}. Each box aggregates results from all weight-sharing combinations $L$.}
    \label{fig:ssl_comparison}
\end{figure}
\subsection{Implementation Details}
\label{sec:implementation_details}
\textbf{AutoPET.} In order to ensure a fair comparison between models, we use the same (811 training + 203 testing) samples of AutoPET \cite{gatidis2022whole} for all models. We utilize Standardized Uptake Values (SUV) of PET scans to reduce inter-patient variation. CT and PET volumes have a voxel size of 2.0mm$\times$2.0mm$\times$3.0mm and the values are clipped to [-100, 250] and [0, 15] for CT and PET respectively. All values are then scaled to [0, 1]. We employ a sliding window inference model by randomly sampling patches of size [96, 96, 96] with a $p=\frac{2}{3}$ probability of containing a tumor. We train using Adam \cite{kingma2014adam} with a learning rate of $10^{-3}$, weight decay of $10^{-6}$, and batch size of 4 for 400 epochs. For Equations \ref{eqn:tf_eq}-\ref{eq:mfc}, we set $\lambda_{\text{rec}}=10^{-4}, \lambda_{\text{seg}}=0.5, \lambda_{\text{class}}=10^{-3}$. Encoder and decoder layers have a kernel size of 3 and stride of 2. Unless otherwise stated, the same data pre-processing is used for comparison to other methods.

\input{tables/all_paradigms}

\textbf{MSD BrainTumor.} We use the same (484 training + 266 testing) samples all models. We train Mirror U-Net using whole MRI volumes with a patch size of [224, 224, 144], with a voxel size of 1.0mm$\times$1.0mm$\times$1.0mm and normalize the intensity for each modality using z-score normalization. We use the same optimization and model hyperparameters as in our AutoPET \cite{gatidis2022whole} experiments.


%% file: tables/notation.tex
\begin{table}[b]
	\scalebox{0.8}{%
			\begin{tabular}{@{}r|l|l}
				\toprule
                    {} & \textbf{Notation} & \textbf{Description} \\ \hline
                     \parbox[t]{2mm}{\multirow{4}{*}{\rotatebox[origin=c]{90}{Input}}} & $(W, H, D) \in \mathbb{N}^3$ & Width, height, and depth of the input volume. \\
				{} & $x \in \mathbb{R}^{W \times H \times D \times 2}$ & PET/CT input volume. \\
                    {} & $x_C \in \mathbb{R}^{W \times H \times D \times 1}$ & CT part of $x$. \\
                    {} & $x_P \in \mathbb{R}^{W \times H \times D \times 1}$ & PET part of $x$. \\ \hline
                     \parbox[t]{2mm}{\multirow{3}{*}{\rotatebox[origin=c]{90}{Output}}} & $y \in \mathbb{R}^{W \times H \times D \times 1}$ & Ground-truth segmentation mask. \\
                    {} & $\hat{y} \in \mathbb{R}^{W \times H \times D \times 1}$ & Predicted segmentation mask. \\
                    {} & $\mathcal{L}_{\text{DiceCE}}(y, \hat{y})$ & Dice Cross-Entropy Loss \cite{isensee2021nnu}. \\ \hline
                     \parbox[t]{2mm}{\multirow{7}{*}{\rotatebox[origin=c]{90}{Model}}} & $E_C(\cdot)$ and $E_P(\cdot)$ & CT- and PET-specific encoder. \\
                    {} & $D_C(\cdot)$ and $D_P(\cdot)$  & CT- and PET-specific decoder. \\
                    {} & $D_{BTL}(\cdot)$ & Bottleneck decoder. \\
                    {} & $T(\cdot)$ & Binary tumor classifier. \\
                    {} & $L \subset \{1...8\}$ & Indices of shared layers. \\
                    {} & $\theta \in [0.1,0.5]$ & Fusion weight for ablation study \textbf{(v4)}. \\
                    {} & $c \in \{0,1\}$ & Binary label -- tumor / no tumor. \\
                    {} & $\oplus$ & Channel concatenation.
 \\ \bottomrule
			\end{tabular}}
   	\caption{Notation used in our equations.}
	\label{tab:notation}
\end{table}

%% file: tables/all_paradigms.tex
\begin{table*}[h!]
    \centering
    \scalebox{0.825}{
    \begin{tabular}{c||c|c|c|c|c|c|c|c}
    \toprule
        {} & {} & \multicolumn{7}{c}{Shared Layers $L$}  \\ \hline
        \multirow{2}{*}{\shortstack{Mirror U-Net \\ \\ Version}}  & \multirow{2}{*}{Parameters} & \multirow{2}{*}{\shortstack{$L=\{3\}$}} & 
        \multirow{2}{*}{\shortstack{$L=\{4\}$}} &
        \multirow{2}{*}{\shortstack{$L=\{5\}$}} &
        \multirow{2}{*}{\shortstack{$L=\{6\}$}} &
        \multirow{2}{*}{\shortstack{$L=\{7\}$}} &
        \multirow{2}{*}{\shortstack{$L = \{4, 5, 6\}$}} &
        \multirow{2}{*}{\shortstack{$L = \{3, 4, 5, 6, 7\}$}}         \\ 
        {} & {} & {} & {} & {} & {} & {} & {} & {} \\ \cline{1-9}

        \multirow{3}{*}{\shortstack{\textbf{(v1)}}} & L2 & 57.39 & 62.88 & 61.92 & 61.91 & 62.06 & 62.08 & 58.06 \\ 
        {} & L2 + noise & 59.76 & 62.83 & 62.34 & 62.21 & 62.65 & 62.36 & 60.86 \\ 
        {} & L2 + shuffling & \textbf{62.75} & \textbf{63.97} & \underline{\textbf{64.57}} & \textbf{63.80} & \textbf{63.09} & \textbf{63.01} & \textbf{62.72} \\ \hline 
        \multirow{3}{*}{\shortstack{\textbf{(v2)}}} & L2 & 58.42 & 63.34 & 62.15 & 61.74 & 62.24 & 62.49 & 59.00 \\ 
        {} & L2 + noise & 59.96 & 64.00 & 62.62 & 62.21 & 62.85 & 63.01 & 61.44 \\ 
        {} & L2 + shuffling & \textbf{63.15} & \textbf{64.28} & \underline{\textbf{65.50}} & \textbf{64.08} & \textbf{63.69} & \textbf{63.29} & \textbf{63.33} \\ \hline  
        \multirow{3}{*}{\shortstack{\textbf{(v3)}}} & L2 & 59.12 & 63.36 & 62.78 & 62.25 & 63.01 & 63.22 & 60.01 \\ 
        {} & L2 + noise & 60.23 & 64.33 & 63.01 & 62.89 & 63.22 & 63.30 & 60.51 \\ 
        {} & L2 + shuffling & \textbf{63.33} & \textbf{64.55} & \underline{\textbf{65.91}} & \textbf{64.44} & \textbf{64.00} & \textbf{63.43} & \textbf{63.34}\\ \hline
        \multirow{6}{*}{\shortstack{\textbf{(v4)}}} & $\theta=0.1$ & 59.61 & \textbf{63.78} & 63.65 & 63.23 & 60.58 & 61.79 & \textbf{60.81} \\ 
        {} & $\theta=0.2$ & 58.65 & 62.66 & \underline{\textbf{64.24}} & 63.35 & \textbf{62.30} & \textbf{63.99} & 56.79 \\  
        {} & $\theta=0.3$ & \textbf{61.67} & 63.76 & 64.22 & 62.26 & 61.89 & 61.33 & 58.21 \\  
        {} & $\theta=0.4$ & 58.95 & 62.93 & 58.43 & \textbf{63.64} & 56.15 & 60.69 & 59.40 \\  
        {} & $\theta=0.5$ & 59.86 & 61.84 & 61.00 & 62.48 & 60.38 & 60.51 & 57.79 \\  
        {} & Learnable $\theta$ & 60.89 & 60.94 & 60.81 & 58.48 & 56.52 & 59.63 & 47.42 \\ 
    \bottomrule
        
    \end{tabular}}
    \caption{Comparison between all Mirror U-Net versions  \textbf{(v1) -- (v4)} for all weight-sharing variants $L$. The best Dice score in each box is in \textbf{bold}. The best Dice score for each Mirror U-Net version \textbf{(v1) -- (v4)} is \underline{underlined}.}
    \label{tab:all_paradigms}
\end{table*}

%% file: sections/experiments.tex
\section{Experiments and Results}
\label{sec:experiments}


\subsection{Task Combinations and Weight Sharing}
\underline{\textbf{Quantitative Results.}} Table \ref{tab:all_paradigms} presents the results on AutoPET \cite{gatidis2022whole} for all Mirror U-Net versions \textbf{(v1)--(v4)} and all weight-sharing variants $L$. We observe three tendencies:

\textbf{(1) Self-Supervision.} Firstly, regularizing the reconstruction by adding noise or voxel shuffling in \textbf{(v1)--(v3)} leads to a consistent improvement regardless of the shared layers $L$, where voxel shuffling achieves the best results. Figure \ref{fig:ssl_comparison} also confirms this tendency and shuffling shows the lowest sensitivity to changes in shared layers $L$, indicated by the lower variance in the box plot, making it not only the best performing but also the most robust method.

\begin{figure}[b]
    \centering
    \includegraphics[scale=0.15]{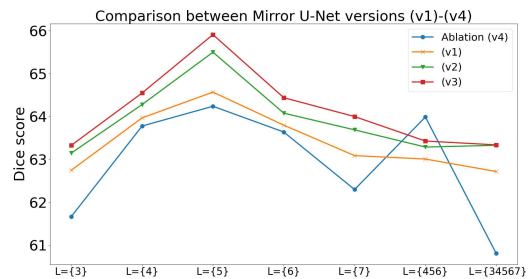}
    \caption{Comparison between all Mirror U-Net versions. The x-axis represents the different weight-sharing schemes. The dots represent the best-performing models for each pair of multi-task settings and weight-sharing scheme, i.e., bold numbers in Table \ref{tab:all_paradigms}.}
    \label{fig:multi-task_comparison}
\end{figure}

\textbf{(2) Weight Sharing.} Secondly, sharing only the bottleneck layer ($L=\{5\})$ results in the best results for all multi-task settings. Figure \ref{fig:multi-task_comparison} shows that the performance behaves similarly for all \textbf{(v1) -- (v4)} when varying shared layers $L$. Sharing shallower, deeper, or multiple layers decreases the performance significantly. The reason for this may be that shallow layers are more modality-specific and deep layers are more task-specific. Sharing such layers does not allow the network to specialize on either the modality or task.

\input{tables/baseline_comparison}

\textbf{(3) Adding Tasks.} Thirdly, Figure \ref{fig:multi-task_comparison} and Table \ref{tab:all_paradigms} show that each Mirror U-Net version consistently outperforms the last \textbf{(v4) $<$ (v1) $<$ (v2) $<$ (v3)}, with the exception of one \textbf{(v4)} outlier. The multi-task settings \textbf{(v1) -- (v3)} share a similar design and build on top of each other by adding new tasks. Hence, incrementally adding meaningful tasks to Mirror U-Net leads to a consistent improvement, regardless of the shared layers $L$.

\begin{figure}[b]
    \centering
    \includegraphics[scale=0.15]{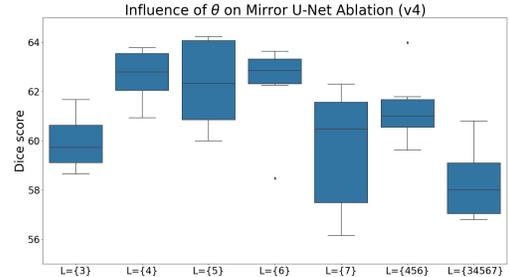}
    \caption{Influence of the fusion parameter $\theta$ on the different weight-sharing schemes for the Decision Fusion setting. Each box aggregates results for $\theta \in [0.1, 0.5]$.}
    \label{fig:decision_fusion_sensitivity}
\end{figure}

\textbf{Ablation (v4) Results.}
The results in Table \ref{tab:all_paradigms} show that the optimal fusion parameter $\theta$ strongly varies for different shared layers. We also train models by setting $\theta$ as a learnable parameter to avoid manual tuning. However, the best scores are achieved with a fixed threshold. Figure \ref{fig:decision_fusion_sensitivity} shows that sharing layers near the bottleneck ($L=\{5\}$) leads to both a higher average performance and a lower sensitivity to changes in $\theta$. Sharing more than one layer leads to a sharp performance loss.  This observation is consistent with the findings in Figure \ref{fig:multi-task_comparison}, which further confirms that sharing only the bottleneck improves not only the performance but also the robustness to parameter changes.

\begin{figure*}[t]
    \centering
    \includegraphics[scale=0.18]{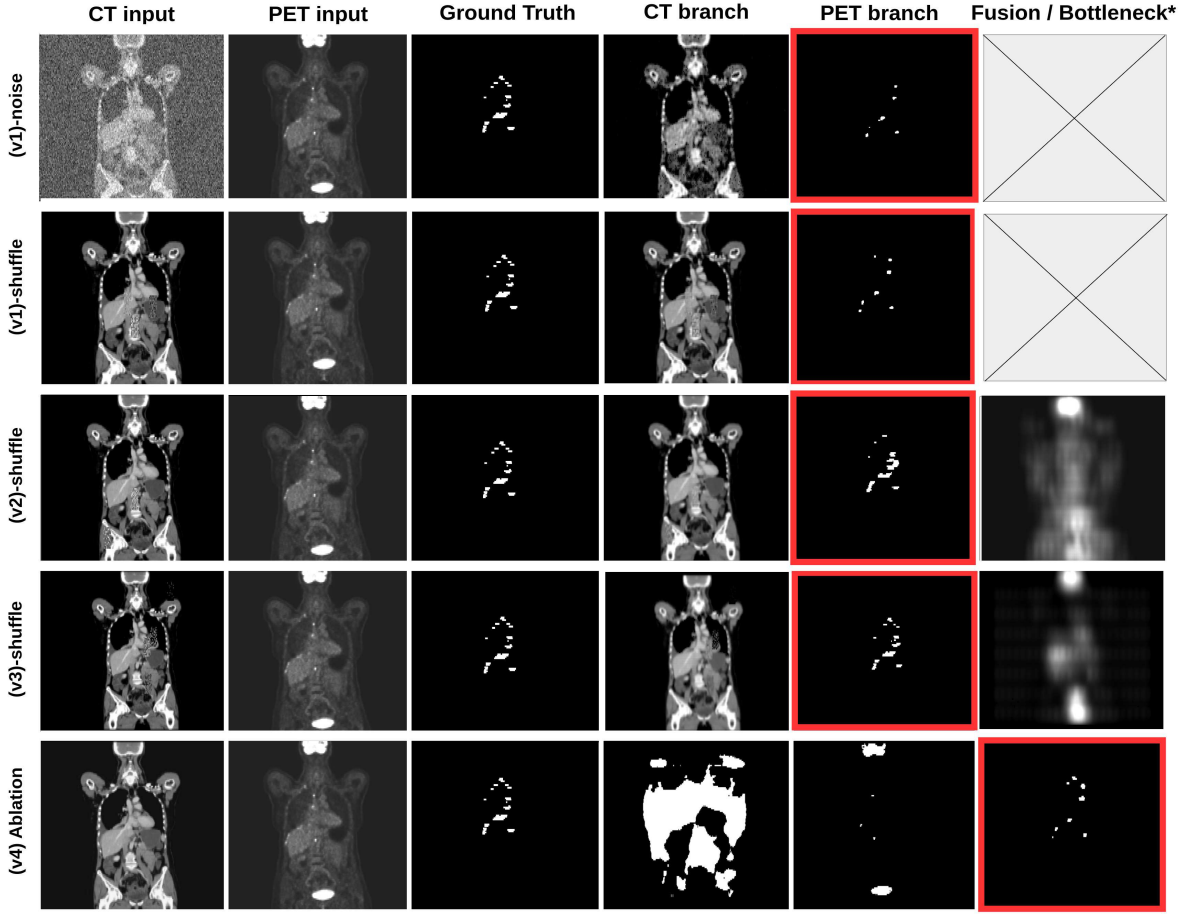}
    \caption{Qualitative results from our multi-task experiments. The last column refers to either the $\theta$-combination of the CT and PET branches from \textbf{(v4)} or to the output of the bottleneck decoder $D_{BTL}$ in \textbf{(v2), (v3)}. The final segmentation for each version is in a red box.}
    \label{fig:qualitative_results}
\end{figure*}

\underline{\textbf{Qualitative Results.}} Figure \ref{fig:qualitative_results} presents the qualitative results for each branch of Mirror U-Net \textbf{(v1) -- (v4)} and gives insight into what each branch has learned.

\textbf{Versions (v1) -- (v3).} As voxel shuffling is the best self-supervision strategy we only include one example for Gaussian noise supervision in \textbf{(v1)-noise}. In the \textbf{(v1)-noise} row, we observe that the CT branch successfully reduces a significant portion of the noise, but struggles with black voxels within the body. On the other hand, in the \textbf{(v1)-shuffle} row, Mirror U-Net restores some edges from the shuffled voxels, which requires the model to remember the structures present within the CT. This leads to a better representation and segmentation results. Moreover, in the \textbf{(v2)-shuffle} model, we reconstruct the PET data from the bottleneck, which is much coarser due to the lack of skip connections. Despite this, the reconstruction includes the regions with the highest metabolic activity, resulting in significantly better segmentation. Finally, in \textbf{(v3)-shuffle}, we add a classification head and the PET reconstruction contains only three regions: the brain, bladder, and the spatial location of the lesions, resulting in much finer boundaries compared to \textbf{(v2)-shuffle}. This refinement leads to a more accurate spatial attention encoded in the bottleneck, as evidenced by the well-delineated lesions in the final segmentation.


\textbf{Ablation (v4).} Since lesion boundaries can be hard to discern in CT scans, the CT branch in \textbf{(v4)} produces a segmentation mask that covers a large portion of the body. However, this mask excludes organs with high metabolic activity, such as the brain, liver, heart, and urinary bladder. On the other hand, the PET branch captures regions with high metabolic activity, including the brain and bladder. Combining the logits from the final layers of each branch results in a fused prediction that better matches the ground-truth mask. These results suggest that the CT branch is not well-suited for segmentation, however, Mirror U-Net's simple architecture allows the CT branch to provide spatial guidance for the PET branch by highlighting regions that are likely to contain lesions and filtering unlikely regions, such as the brain and bladder.

Overall, our qualitative results demonstrate that Mirror U-Net utilizes the complementary nature of CT and PET images and successfully transfers knowledge via the shared features to improve the primary segmentation task.

\subsection{Comparisons to Other Approaches}
\textbf{Comparison to Baselines.}
We compare Mirror U-Net to several baselines -- traditional early (EF), middle (MF), and late fusion (LF), as well as a unimodal U-Net \cite{ronneberger2015u} trained solely on either CT or PET data. We compare to three late fusion variants: sum of logits (LF-Logits), and predictions' union (LF-$\cup$) or intersection (LF-$\cap$). We use the Dice score and the average false positive (FPV) and false negative volumes (FNV) as evaluation metrics, which measure the average volume of over- (FPV) and under-segmented (FPN) regions in mm$^3$. The results, presented in Table \ref{tab:baseline_comparisons}, show that the unimodal CT model has poor performance, as lesions are hardly visible in CT scans, which limits its potential as a standalone modality. The unimodal PET model, on the other hand, outperforms all traditional fusion strategies, highlighting the limitations of traditional fusion in effectively combining PET and CT features without extensive fine-tuning. However, the PET model has a higher FPV since it is challenging to distinguish lesions from highly active organs based solely on PET data. The union late fusion achieves the highest Dice score among the fusion baselines, whereas the intersection late fusion has the lowest FPV due to the limited agreement between PET and CT predictions. All variants of Mirror U-Net consistently outperform the baseline methods on all metrics.

\input{tables/comparison_sota}

\textbf{Comparison to Related Work.}
Our main contribution is the combination of multimodal fission and multi-task learning. Therefore, we compare Mirror U-Net to related methods that utilize only fission, only multi-task learning, or neither approach. For the \textbf{neither} category, we compare Mirror U-Net to nnUNet \cite{isensee2021nnu}, a standard benchmark for medical segmentation. Additionally, we include Blackbean \cite{ye2022exploring}, the winner of the AutoPET 2022 challenge, to demonstrate state-of-the-art performance on AutoPET \cite{gatidis2022whole}.

In the \textbf{multi-task-only} category, we compare to related approaches using multi-task learning and multimodal \textbf{fusion}. SF-Net \cite{liu2022sf} utilizes a decoder branch to reconstruct an image fusion of T1c and T2 MRI while preserving the structures of both modalities via an L2 and SSIM loss \cite{zeng20123d}. Andrearczyk \etal \cite{andrearczyk2021multi} and DeepMTS \cite{meng2022deepmts} utilize feature maps from a U-Net model \cite{ronneberger2015u} to train a classifier for survival prediction. Weninger \etal \cite{weninger2019multi} utilize reconstruction and tumor classification (enhancing or non-enhancing) to regularize an early fusion U-Net model \cite{ronneberger2015u}. Lastly, we conduct an ablation study where we only use either CT or PET-only data in all branches in \textbf{(v1)-(v3)} to show that the presence of both modalities is necessary for Mirror U-Net.

\begin{figure}[b]
    \centering
    \includegraphics[width=\linewidth]{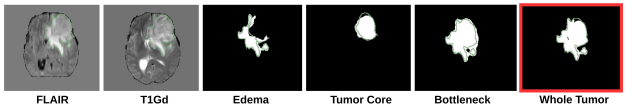}
    \caption{Qualitative results for Mirror U-Net \textbf{(v2)} on brain tumor segmentation. The final whole tumor prediction is in a red box and is obtained by the union of the edema and tumor core predictions.}
    \label{fig:brats_qualitative}
\end{figure}

For the \textbf{fission-only category} We compare to the single-task model of Valindria \etal \cite{valindria2018multi} which uses a similar architecture to Mirror U-Net, but alternates between modalities during each iteration and uses both branches for segmentation. We also consider Mirror U-Net \textbf{(v4)} as a fission-only method as it only has one segmentation task.

We train and evaluate all models on the same 80/20 training/validation split and use the same data preprocessing as Mirror U-Net, except for nnUNet \cite{isensee2021nnu} and Blackbean \cite{ye2022exploring}, which require specific preprocessing steps. The results in Table \ref{tab:sota_comparisons} show that Mirror U-Net consistently outperforms all other models, demonstrating state-of-the-art performance on AutoPET \cite{gatidis2022whole}. This underscores the power of combining multimodal fission with multi-task learning, highlighting the efficacy of our proposed method.

\subsection{Generalizing to Brain Tumor Segmentation}
We compare Mirror U-Net \textbf{(v2)} to nnUNet \cite{isensee2021nnu}, SegResNet \cite{myronenko20193d}, which is the winner of the Brain Tumor Segmentation BraTS 2018 challenge, and to Mirror U-Net \textbf{(v1)} and \textbf{(v2)-rec} as ablations. Our findings, shown in Table \ref{tab:brats}, demonstrate that Mirror U-Net outperforms all other methods on all 3 tumor classes. Notably, we observe a significant performance drop when the bottleneck task is omitted in Mirror U-Net \textbf{(v1)}. Using the default tasks in \textbf{(v2)-rec} achieves the second-best results for tumor core and edema segmentation. However, unlike CT, both FLAIR and T1Gd modalities have a strong signal, making segmentation tasks more suitable than the default reconstruction.

In Figure \ref{fig:brats_qualitative}, we show that the bottleneck layer has learned a coarse segmentation mask of the whole tumor, with some over-segmentation. However, this coarse segmentation mask provides valuable spatial guidance for the edema and core segmentation tasks, resulting in a much finer final segmentation of the whole tumor. These qualitative results, combined with our quantitative findings in Table \ref{tab:brats}, suggest that Mirror U-Net has the potential to generalize well to other imaging modalities and tasks beyond the specific PET/CT segmentation task studied in this work.

\input{tables/brats}

%% file: tables/baseline_comparison.tex
\begin{table*}[!h]
    \centering
    \scalebox{0.875}{
    \begin{tabular}{r|ccccccc|ccccc}
    \toprule
        Metric & \multicolumn{7}{c|}{Baselines} & \multicolumn{4}{c}{Mirror U-Net (Ours)} \\ \hline
        {} & CT & PET & EF & MF & LF-Logit & LF-$\cup$ & LF-$\cap$   & \textbf{(v1)} & \textbf{(v2)}  & \textbf{(v3)} & Ablation \textbf{(v4)}  \\ \hline
        Dice $\uparrow$ & 26.00 & \textbf{60.99} & 54.89 & 55.53 & 57.41 & 59.89 & 21.60 & 64.57 & 65.50 & \textbf{65.91} & 64.24 \\ 
        FPV $\downarrow$ & 15.64 & 5.38 & 4.98 & 4.77 &4.88 & 3.95 & \textbf{1.67} & 2.93 & 2.83 & \textbf{1.55} & 2.93 \\ 
        FNV $\downarrow$ & 44.15 & \textbf{2.15} & 3.13 & 3.02 & 2.88 & 3.01 & 99.74 & 1.66 & 0.94 & \textbf{0.76} & 1.99 \\ \bottomrule
    \end{tabular}}
    \caption{Comparison to the baselines with the same U-Net backbone as Mirror U-Net. EF: Early, MF: Middle, LF: Late Fusion.}
    \label{tab:baseline_comparisons}
\end{table*}

%% file: tables/comparison_sota.tex
\begin{table*}[t]
    \centering
    \scalebox{0.8}{
    \begin{tabular}{l||c|c|c|c|c|c}
    \toprule
        Method & Dice $\uparrow$ & FPV $\downarrow$ & FNV $\downarrow$  & Tasks & Multimodal Fission & Multi-task \\ \hline
        nnUNet \cite{isensee2021nnu} & 62.75 & 2.83 & 1.59 & Seg &  {} & {} \\ 
        Blackbean \cite{ye2022exploring} & 63.15 & 2.55 & 1.76 & Seg & {} & {} \\ \hline
        SF-Net \cite{liu2022sf} & 61.21 & 3.44 & 2.95 & Seg + Rec & {} & \checkmark \\ 
        Andrearczyk et al. \cite{andrearczyk2021multi} & 61.45 & 2.98 & 1.89 & Seg + Class & {} & \checkmark \\

        DeepMTS \cite{meng2022deepmts} & 61.91 & 3.22 & 2.76 & Seg + Class & {} & \checkmark \\ 

        Weninger et al. \cite{weninger2019multi} & 61.22 & 3.98 & 2.82 & Seg + Rec + Class & {} & \checkmark \\ 
        CT-only Mirror U-Net \textbf{(v3)} & 12.37 & 28.24 & 50.02 & Seg + Rec + Class & {} & \checkmark \\
        PET-only Mirror U-Net \textbf{(v3)} & 56.14 & 4.81 & 3.02 & Seg + Rec + Class & {} & \checkmark \\ \hline
        Mirror U-Net \textbf{(v4)} & 64.24 & 2.93 & 1.99 & Seg & \checkmark & {} \\
        Valindria \etal \cite{valindria2018multi} & 39.84 & 7.89 & 17.00 & Seg & \checkmark & {}   \\ \hline
        (Ours) Mirror U-Net \textbf{(v3)} & \textbf{65.91} & \textbf{1.55} & \textbf{0.76} & Seg + Rec + Class & \checkmark & \checkmark \\

    \bottomrule
    \end{tabular}}
    \caption{Comparison to related fission-only and multi-task only methods and to the current state-of-the-art on the AutoPET dataset \cite{gatidis2022whole, ye2022exploring}.}
    \label{tab:sota_comparisons}
\end{table*}

%% file: tables/brats.tex
\begin{table}[!h]
    \centering
    \scalebox{0.75}{
    \begin{tabular}{l||c|c|c}
    \toprule
        {} & \multicolumn{3}{c}{Dice $\uparrow$} \\ \hline
        Method & Whole Tumor & Edema & Tumor Core  \\ \hline
        Mirror U-Net \textbf{(v1)} & 88.37 & 71.91 & 82.22 \\
        Mirror U-Net \textbf{(v2)-rec} & 91.12 & \underline{77.12} & \underline{84.56} \\

        nnUNet \cite{isensee2021nnu} & 89.10 & 76.35 & 84.05 \\ 
        SegResNet \cite{myronenko20193d} & \underline{91.29} & 77.01 & 84.22  \\ \hline
        (Ours) Mirror U-Net \textbf{(v2)} & \textbf{92.52} & \textbf{78.12} & \textbf{85.84} \\ 
    \bottomrule
    \end{tabular}}
    \caption{Comparison to other methods on MSD BrainTumor \cite{antonelli2022medical}.}
    \label{tab:brats}
\end{table}

%% file: sections/conclusion.tex
\section{Conclusion and Discussion}
\label{sec:conclusion}

In summary, we propose Mirror U-Net, which combines multimodal fission and multi-task learning for the first time. Our model outperforms traditional fusion methods as well as fission-only and multi-task-only approaches on the AutoPET 2022 Challenge and shows state-of-the-art performance, which demonstrates the power of combining fission with multi-task learning. Our results indicate that sharing only the bottleneck layer is optimal while sharing shallower or deeper layers leads to a performance drop. We also demonstrate the generalizability of Mirror U-Net to brain tumor segmentation from multimodal MRI scans. Our qualitative experiments reveal that selecting appropriate tasks improves performance as the shared representation learns a spatial guidance that boosts the primary segmentation task. Our model's robustness to hyperparameter changes and high performance is a promising step toward deploying PET/CT segmentation models in clinical practice.


\section{Acknowledgements}
The present contribution is supported by the Helmholtz Association under the joint research school “HIDSS4Health – Helmholtz Information and Data Science School for Health. This work was performed on the HoreKa supercomputer funded by the Ministry of Science, Research and the Arts Baden-Württemberg and by
the Federal Ministry of Education and Research.

%% file: egpaper_final.bbl
\begin{thebibliography}{10}\itemsep=-1pt

\bibitem{andrearczyk2021multi}
Vincent Andrearczyk, Pierre Fontaine, Valentin Oreiller, Joel Castelli, Mario
  Jreige, John~O Prior, and Adrien Depeursinge.
\newblock Multi-task deep segmentation and radiomics for automatic prognosis in
  head and neck cancer.
\newblock In {\em International Workshop on PRedictive Intelligence In
  MEdicine}, pages 147--156. Springer, 2021.

\bibitem{antonelli2022medical}
Michela Antonelli, Annika Reinke, Spyridon Bakas, Keyvan Farahani, Annette
  Kopp-Schneider, Bennett~A Landman, Geert Litjens, Bjoern Menze, Olaf
  Ronneberger, Ronald~M Summers, et~al.
\newblock The medical segmentation decathlon.
\newblock {\em Nature communications}, 13(1):4128, 2022.

\bibitem{ben200918f}
Simona Ben-Haim and Peter Ell.
\newblock 18f-fdg pet and pet/ct in the evaluation of cancer treatment
  response.
\newblock {\em Journal of Nuclear Medicine}, 50(1):88--99, 2009.

\bibitem{bendazzoli2022priornet}
Simone Bendazzoli and Mehdi Astaraki.
\newblock Priornet: lesion segmentation in pet-ct including prior tumor
  appearance information.
\newblock {\em arXiv preprint arXiv:2210.02203}, 2022.

\bibitem{bi2014multi}
Lei Bi, Jinman Kim, Dagan Feng, and Michael Fulham.
\newblock Multi-stage thresholded region classification for whole-body pet-ct
  lymphoma studies.
\newblock In {\em International Conference on Medical Image Computing and
  Computer-Assisted Intervention}, pages 569--576. Springer, 2014.

\bibitem{bourigault2021multimodal}
Emmanuelle Bourigault, Daniel~R McGowan, Abolfazl Mehranian, and
  Bart{\l}omiej~W Papie{\.z}.
\newblock Multimodal pet/ct tumour segmentation and prediction of
  progression-free survival using a full-scale unet with attention.
\newblock In {\em 3D Head and Neck Tumor Segmentation in PET/CT Challenge},
  pages 189--201. Springer, 2021.

\bibitem{cheng2022fully}
Jianhong Cheng, Jin Liu, Hulin Kuang, and Jianxin Wang.
\newblock A fully automated multimodal mri-based multi-task learning for glioma
  segmentation and idh genotyping.
\newblock {\em IEEE Transactions on Medical Imaging}, 2022.

\bibitem{cciccek20163d}
{\"O}zg{\"u}n {\c{C}}i{\c{c}}ek, Ahmed Abdulkadir, Soeren~S Lienkamp, Thomas
  Brox, and Olaf Ronneberger.
\newblock 3d u-net: learning dense volumetric segmentation from sparse
  annotation.
\newblock In {\em International conference on medical image computing and
  computer-assisted intervention}, pages 424--432. Springer, 2016.

\bibitem{fu2021multimodal}
Xiaohang Fu, Lei Bi, Ashnil Kumar, Michael Fulham, and Jinman Kim.
\newblock Multimodal spatial attention module for targeting multimodal pet-ct
  lung tumor segmentation.
\newblock {\em IEEE Journal of Biomedical and Health Informatics},
  25(9):3507--3516, 2021.

\bibitem{gatidis2022whole}
Sergios Gatidis, Tobias Hepp, Marcel Fr{\"u}h, Christian La~Foug{\`e}re,
  Konstantin Nikolaou, Christina Pfannenberg, Bernhard Sch{\"o}lkopf, Thomas
  K{\"u}stner, Clemens Cyran, and Daniel Rubin.
\newblock A whole-body fdg-pet/ct dataset with manually annotated tumor
  lesions.
\newblock {\em Scientific Data}, 9(1):1--7, 2022.

\bibitem{guo2019deep}
Zhe Guo, Xiang Li, Heng Huang, Ning Guo, and Quanzheng Li.
\newblock Deep learning-based image segmentation on multimodal medical imaging.
\newblock {\em IEEE Transactions on Radiation and Plasma Medical Sciences},
  3(2):162--169, 2019.

\bibitem{hallitschke2023multimodal}
Verena~Jasmin Hallitschke, Tobias Schlumberger, Philipp Kataliakos, Zdravko
  Marinov, Moon Kim, Lars Heiliger, Constantin Seibold, Jens Kleesiek, and
  Rainer Stiefelhagen.
\newblock Multimodal interactive lung lesion segmentation: A framework for
  annotating pet/ct images based on physiological and anatomical cues.
\newblock {\em arXiv preprint arXiv:2301.09914}, 2023.

\bibitem{heiliger2022autopet}
Lars Heiliger, Zdravko Marinov, Max Hasin, Andr{\'e} Ferreira, Jana Fragemann,
  Kelsey Pomykala, Jacob Murray, David Kersting, Victor Alves, Rainer
  Stiefelhagen, et~al.
\newblock Autopet challenge: Combining nn-unet with swin unetr augmented by
  maximum intensity projection classifier.
\newblock {\em arXiv preprint arXiv:2209.01112}, 2022.

\bibitem{hickson2022sharing}
Steven Hickson, Karthik Raveendran, and Irfan Essa.
\newblock Sharing decoders: Network fission for multi-task pixel prediction.
\newblock In {\em Proceedings of the IEEE/CVF Winter Conference on Applications
  of Computer Vision}, pages 3771--3780, 2022.

\bibitem{hickson2019floors}
Steven Hickson, Karthik Raveendran, Alireza Fathi, Kevin Murphy, and Irfan
  Essa.
\newblock Floors are flat: Leveraging semantics for real-time surface normal
  prediction.
\newblock In {\em Proceedings of the IEEE/CVF International Conference on
  Computer Vision Workshops}, pages 0--0, 2019.

\bibitem{hsu2018disentangling}
Wei-Ning Hsu and James Glass.
\newblock Disentangling by partitioning: A representation learning framework
  for multimodal sensory data.
\newblock {\em arXiv preprint arXiv:1805.11264}, 2018.

\bibitem{isensee2021nnu}
Fabian Isensee, Paul~F Jaeger, Simon~AA Kohl, Jens Petersen, and Klaus~H
  Maier-Hein.
\newblock nnu-net: a self-configuring method for deep learning-based biomedical
  image segmentation.
\newblock {\em Nature methods}, 18(2):203--211, 2021.

\bibitem{isensee2017brain}
Fabian Isensee, Philipp Kickingereder, Wolfgang Wick, Martin Bendszus, and
  Klaus~H Maier-Hein.
\newblock Brain tumor segmentation and radiomics survival prediction:
  Contribution to the brats 2017 challenge.
\newblock In {\em International MICCAI Brainlesion Workshop}, pages 287--297.
  Springer, 2017.

\bibitem{jiang2020self}
Jue Jiang, Yu-Chi Hu, Neelam Tyagi, Chuang Wang, Nancy Lee, Joseph~O Deasy,
  Berry Sean, and Harini Veeraraghavan.
\newblock Self-derived organ attention for unpaired ct-mri deep domain
  adaptation based mri segmentation.
\newblock {\em Physics in Medicine \& Biology}, 65(20):205001, 2020.

\bibitem{jiang2019cross}
Jue Jiang, Yu-Chi Hu, Neelam Tyagi, Pengpeng Zhang, Andreas Rimner, Joseph~O
  Deasy, and Harini Veeraraghavan.
\newblock Cross-modality (ct-mri) prior augmented deep learning for robust lung
  tumor segmentation from small mr datasets.
\newblock {\em Medical physics}, 46(10):4392--4404, 2019.

\bibitem{jiang2019two}
Zeyu Jiang, Changxing Ding, Minfeng Liu, and Dacheng Tao.
\newblock Two-stage cascaded u-net: 1st place solution to brats challenge 2019
  segmentation task.
\newblock In {\em International MICCAI brainlesion workshop}, pages 231--241.
  Springer, 2019.

\bibitem{joze2020mmtm}
Hamid Reza~Vaezi Joze, Amirreza Shaban, Michael~L Iuzzolino, and Kazuhito
  Koishida.
\newblock Mmtm: Multimodal transfer module for cnn fusion.
\newblock In {\em Proceedings of the IEEE/CVF Conference on Computer Vision and
  Pattern Recognition}, pages 13289--13299, 2020.

\bibitem{kendall2018multi}
Alex Kendall, Yarin Gal, and Roberto Cipolla.
\newblock Multi-task learning using uncertainty to weigh losses for scene
  geometry and semantics.
\newblock In {\em Proceedings of the IEEE conference on computer vision and
  pattern recognition}, pages 7482--7491, 2018.

\bibitem{kingma2014adam}
Diederik~P Kingma and Jimmy Ba.
\newblock Adam: A method for stochastic optimization.
\newblock {\em arXiv preprint arXiv:1412.6980}, 2014.

\bibitem{kuga2017multi}
Ryohei Kuga, Asako Kanezaki, Masaki Samejima, Yusuke Sugano, and Yasuyuki
  Matsushita.
\newblock Multi-task learning using multi-modal encoder-decoder networks with
  shared skip connections.
\newblock In {\em Proceedings of the IEEE International Conference on Computer
  Vision Workshops}, pages 403--411, 2017.

\bibitem{liang2022foundations}
Paul~Pu Liang, Amir Zadeh, and Louis-Philippe Morency.
\newblock Foundations and recent trends in multimodal machine learning:
  Principles, challenges, and open questions.
\newblock {\em arXiv preprint arXiv:2209.03430}, 2022.

\bibitem{liu2022sf}
Yu Liu, Fuhao Mu, Yu Shi, and Xun Chen.
\newblock Sf-net: A multi-task model for brain tumor segmentation in multimodal
  mri via image fusion.
\newblock {\em IEEE Signal Processing Letters}, 29:1799--1803, 2022.

\bibitem{liu2022autopet}
Zhantao Liu, Shaonan Zhong, and Junyang Mo.
\newblock Autopet challenge 2022: Step-by-step lesion segmentation in
  whole-body fdg-pet/ct.
\newblock {\em arXiv preprint arXiv:2209.09199}, 2022.

\bibitem{maninis2019attentive}
Kevis-Kokitsi Maninis, Ilija Radosavovic, and Iasonas Kokkinos.
\newblock Attentive single-tasking of multiple tasks.
\newblock In {\em Proceedings of the IEEE/CVF Conference on Computer Vision and
  Pattern Recognition}, pages 1851--1860, 2019.

\bibitem{marinov2022modselect}
Zdravko Marinov, Alina Roitberg, David Schneider, and Rainer Stiefelhagen.
\newblock Modselect: Automatic modality selection for synthetic-to-real domain
  generalization.
\newblock {\em arXiv preprint arXiv:2208.09414}, 2022.

\bibitem{meng2022deepmts}
Mingyuan Meng, Bingxin Gu, Lei Bi, Shaoli Song, David~Dagan Feng, and Jinman
  Kim.
\newblock Deepmts: Deep multi-task learning for survival prediction in patients
  with advanced nasopharyngeal carcinoma using pretreatment pet/ct.
\newblock {\em IEEE Journal of Biomedical and Health Informatics},
  26(9):4497--4507, 2022.

\bibitem{menze2014multimodal}
Bjoern~H Menze, Andras Jakab, Stefan Bauer, Jayashree Kalpathy-Cramer, Keyvan
  Farahani, Justin Kirby, Yuliya Burren, Nicole Porz, Johannes Slotboom, Roland
  Wiest, et~al.
\newblock The multimodal brain tumor image segmentation benchmark (brats).
\newblock {\em IEEE transactions on medical imaging}, 34(10):1993--2024, 2014.

\bibitem{milletari2016v}
Fausto Milletari, Nassir Navab, and Seyed-Ahmad Ahmadi.
\newblock V-net: Fully convolutional neural networks for volumetric medical
  image segmentation.
\newblock In {\em 2016 fourth international conference on 3D vision (3DV)},
  pages 565--571. IEEE, 2016.

\bibitem{mlynarski2019deep}
Pawel Mlynarski, Herv{\'e} Delingette, Antonio Criminisi, and Nicholas Ayache.
\newblock Deep learning with mixed supervision for brain tumor segmentation.
\newblock {\em Journal of Medical Imaging}, 6(3):034002, 2019.

\bibitem{myronenko20193d}
Andriy Myronenko.
\newblock 3d mri brain tumor segmentation using autoencoder regularization.
\newblock In {\em Brainlesion: Glioma, Multiple Sclerosis, Stroke and Traumatic
  Brain Injuries: 4th International Workshop, BrainLes 2018, Held in
  Conjunction with MICCAI 2018, Granada, Spain, September 16, 2018, Revised
  Selected Papers, Part II 4}, pages 311--320. Springer, 2019.

\bibitem{peng2022automatic}
Yige Peng, Jinman Kim, Dagan Feng, and Lei Bi.
\newblock Automatic tumor segmentation via false positive reduction network for
  whole-body multi-modal pet/ct images.
\newblock {\em arXiv preprint arXiv:2209.07705}, 2022.

\bibitem{ronneberger2015u}
Olaf Ronneberger, Philipp Fischer, and Thomas Brox.
\newblock U-net: Convolutional networks for biomedical image segmentation.
\newblock In {\em International Conference on Medical image computing and
  computer-assisted intervention}, pages 234--241. Springer, 2015.

\bibitem{shu2022expansion}
Xiangbo Shu, Jiawen Yang, Rui Yan, and Yan Song.
\newblock Expansion-squeeze-excitation fusion network for elderly activity
  recognition.
\newblock {\em IEEE Transactions on Circuits and Systems for Video Technology},
  2022.

\bibitem{sibille2022whole}
Ludovic Sibille, Xinrui Zhan, and Lei Xiang.
\newblock Whole-body tumor segmentation of 18f-fdg pet/ct using a cascaded and
  ensembled convolutional neural networks.
\newblock {\em arXiv preprint arXiv:2210.08068}, 2022.

\bibitem{townsend2004pet}
David~W Townsend, Jonathan~PJ Carney, Jeffrey~T Yap, and Nathan~C Hall.
\newblock Pet/ct today and tomorrow.
\newblock {\em Journal of Nuclear Medicine}, 45(1 suppl):4S--14S, 2004.

\bibitem{tsai2018learning}
Yao-Hung~Hubert Tsai, Paul~Pu Liang, Amir Zadeh, Louis-Philippe Morency, and
  Ruslan Salakhutdinov.
\newblock Learning factorized multimodal representations.
\newblock In {\em International Conference on Learning Representations}, 2018.

\bibitem{valindria2018multi}
Vanya~V Valindria, Nick Pawlowski, Martin Rajchl, Ioannis Lavdas, Eric~O
  Aboagye, Andrea~G Rockall, Daniel Rueckert, and Ben Glocker.
\newblock Multi-modal learning from unpaired images: Application to multi-organ
  segmentation in ct and mri.
\newblock In {\em 2018 IEEE winter conference on applications of computer
  vision (WACV)}, pages 547--556. IEEE, 2018.

\bibitem{weisman2022automated}
Amy Weisman, Ojaswita Lokre, Brayden Schott, Victor Fernandes, Robert Jeraj,
  Timothy Perk, Steve Cho, and Scott Perlman.
\newblock Automated detection and quantification of neuroendocrine tumors on
  68ga-dotatate pet/ct images using a u-net ensemble method, 2022.

\bibitem{weninger2019multi}
Leon Weninger, Qianyu Liu, and Dorit Merhof.
\newblock Multi-task learning for brain tumor segmentation.
\newblock In {\em International MICCAI brainlesion workshop}, pages 327--337.
  Springer, 2019.

\bibitem{xue2021multi}
Zhongliang Xue, Ping Li, Liang Zhang, Xiaoyuan Lu, Guangming Zhu, Peiyi Shen,
  Syed Afaq~Ali Shah, and Mohammed Bennamoun.
\newblock Multi-modal co-learning for liver lesion segmentation on pet-ct
  images.
\newblock {\em IEEE Transactions on Medical Imaging}, 40(12):3531--3542, 2021.

\bibitem{yap2004image}
Jeffrey~T Yap, Jonathan~PJ Carney, Nathan~C Hall, and David~W Townsend.
\newblock Image-guided cancer therapy using pet/ct.
\newblock {\em The Cancer Journal}, 10(4):221--233, 2004.

\bibitem{ye2022exploring}
Jin Ye, Haoyu Wang, Ziyan Huang, Zhongying Deng, Yanzhou Su, Can Tu, Qian Wu,
  Yuncheng Yang, Meng Wei, Jingqi Niu, et~al.
\newblock Exploring vanilla u-net for lesion segmentation from whole-body
  fdg-pet/ct scans.
\newblock {\em arXiv preprint arXiv:2210.07490}, 2022.

\bibitem{zeng20123d}
Kai Zeng and Zhou Wang.
\newblock 3d-ssim for video quality assessment.
\newblock In {\em 2012 19th IEEE international conference on image processing},
  pages 621--624. IEEE, 2012.

\bibitem{zhang2022whole}
Jia Zhang, Yukun Huang, Zheng Zhang, and Yuhang Shi.
\newblock Whole-body lesion segmentation in 18f-fdg pet/ct.
\newblock {\em arXiv preprint arXiv:2209.07851}, 2022.

\bibitem{zhang2018translating}
Zizhao Zhang, Lin Yang, and Yefeng Zheng.
\newblock Translating and segmenting multimodal medical volumes with cycle-and
  shape-consistency generative adversarial network.
\newblock In {\em Proceedings of the IEEE conference on computer vision and
  pattern Recognition}, pages 9242--9251, 2018.

\bibitem{zhong2022autopet}
Shaonan Zhong, Junyang Mo, and Zhantao Liu.
\newblock Autopet challenge 2022: Automatic segmentation of whole-body tumor
  lesion based on deep learning and fdg pet/ct.
\newblock {\em arXiv preprint arXiv:2209.01212}, 2022.

\end{thebibliography}
